# Alternative forms of representation of Boolean functions in Cryptographic Information Security Facilities

## Kushch S.

The work offers a new approach to the formation of functions which are used in cryptography and cryptanalysis. It will use alternative forms of representation of Boolean functions, that is, those which are different from the classical form, which is formed in a Boolean basis AND-OR-NOT.

An example of this is, in particular, the formation of cryptographic functions with the use of alternative forms of representation, namely Cognate-representation of Boolean functions. This form is, by definition, multivariant and allows you to choose the best variant from a plurality of possible and permissible forms. Moreover, criteria of admissibility can be also selected depending on the particular situation, since it is known that an improvement of one criterion usually leads to deterioration of others.

The methods suggested in the project, exemplified by Cognate-form representations of Boolean functions, show that the use of alternative forms of representation of Boolean functions in forming of cryptographic functions, algorithms and devices can significantly improve their parameters and properties.

And their use in cryptographic means of protection allow to optimize the process of logical design of cryptographic devices and improve the safety performance of information and communication systems.

*Keywords: cryptography, cryptographic functions, cryptographic properties, forming algorithms cryptographic functions, Cognate-forms of Boolean functions representations, alternative forms of representation of Boolean functions, Cognate implementation, cyber security.*

INTRODUCTION

Prompt development of information technology has a significant influence on all aspects of human life. The time of mass communications, the Internet, informatization management of technological processes in various spheres of human activity have resulted in a sharp rise of providing the security of information systems from unauthorized access and destructive influences. As a result, the tasks of constructing reliable telecommunications systems and the development of methods for assessing their level of protection are becoming of high priority nowadays. Further, the cryptographic protection means are playing the important role in ensuring the security of information transmission. Coincidently, the experience of the practical usage of existing cryptographic means of protecting information shows, that the systems which are applied practically are not always able to provide the modern information security requirements. Therefore the task of improving means of information security is integral and urgent.

The article target is to search for variants to optimize the use of alternative forms of representation of Boolean functions in the Cryptography. We call the alternative forms of representation (FR) of Boolean functions (BF) that differ from the classical FR (basis AND-OR-NOT), the most common currently.

Namely the algebraic FR, which is the result of ortofunktsional F-transformation Boolean functions to the equivalent piecewise constant functions, the Reed-Muller FR, where BF submitted as system of BF in basis AND-SUM MOD2-AND [3] and well researched in recent years, the Orthogonal-FR [7] and Cognate-FR [5,4], etc.

The BF is one of the basic structural elements in the most modern cryptographic structures (stream ciphers, block ciphers, hash functions, etc.). Such functions (system functions), which are used in the synthesis of cryptographic objects are named as cryptographic functions. The list of mathematical requirements (properties) was selected during the development of means and methods of cryptographic analysis which must satisfy cryptographic functions. The existence of such properties in the functions is intended to ensure the resistance of cryptographic schemes to cryptanalysis. Examples of such properties are: the lack of correlation between the value of the function and a set of the variables fixed cardinality [8], the lack in BF the low-degree annihilators [1], the absence of BF (mapping) in linear structures [2]. Sets of BF, with these properties are allocated to the separate classes. These include the bent-function, the correlation-immune functions, the algebraically-immune functions and the algebraic non-degenerate functions. A characteristic feature of these classes is not only the absence of exact algebraic description, but as well as the lack of exact expressions to assess their capacity. The examples of the results of researches in this area may be the works of Maitra for correlation-immune functions, the estimates for the number of bent functions in the works of Carlet and Krotov, the asymptotic estimate of the number of an algebraically degenerate functions. Although similar functions have nontrivial linear structures that do not have the necessary cryptographic properties, however, they have an important role in cryptanalysis.

The set of cryptographic functions may be represented in the form of a Venn diagram (fig.1).

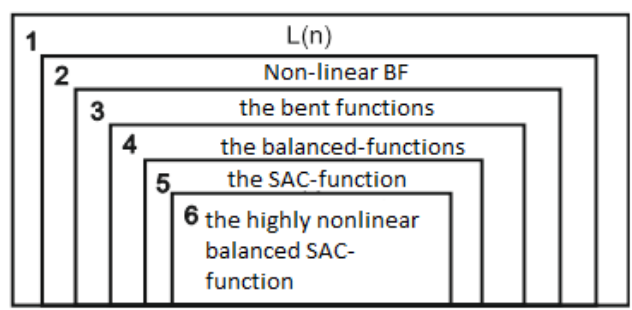

Figure 1. Venn diagram for cryptographic Boolean functions

The study of cryptographic properties occur using different FR of BF, such as the algebraic, the normal form, the numerical normal form, the polynomial representation by expanding of a field of two elements, the presentation with using the graphs and others. In the analysis of the properties of cryptographic functions are used the results of mathematical cybernetics, of combinatorial analysis and algebra. An important role is played by experimental studies with use of opportunities of computer technology.

In order to draw a conclusion about the possibility of using alternative forms of representation of Boolean functions in cryptography and cryptanalysis is necessary to carry out the following:

- research of cryptographic properties and construction of broad classes of BF which have the desired properties and construction of a functions which have different extreme parameters in the different FR;
- research of cryptographic properties of BF implemented in various, including the alternatives FR in the concrete cryptographic systems;
- construction of BF sets which have two or more necessary cryptographic properties, as well as others that may emerge during research.

Although the research of the properties of functions of the concrete cryptographic systems became a typical task of any cryptanalysis, at the same time there are some tasks which fulfillment would allow us to reach a new level of formation of cryptographic systems. These include, for example:

- the calculation of cardinality (or their assessment) of some BF classes, for example, BF which have the property of correlation immunity of a given order, the bent function, the k-bent functions [9] in various FR, etc;
- the description of the group of invariance of the concrete cryptographic properties;
- development of algorithms for the approximation of an arbitrary function by function from a given class.

Taking this into account, the existing set of cryptographic properties of BF and their reflections in a various FR are no way can be considered complete. The practice shows that methods of construction of cryptographic functions and cryptographic analysis continue to develop nowadays, and the results of such development put forward more and more new requirements to the cryptographic functions. That is why the search for new variants of cryptographic functions formation is an important scientific and practical task of the day. In the papers [3-7] is showed and proved that the use of alternative FR allows to simplify the creation and technical implementation of BF. Furthermore, it is also possible to carry out the implementation of the big cryptographic functions of piecemeal. It will be used in the formation difficult cryptographic functions as well as for the implementation of technical means for information protection.

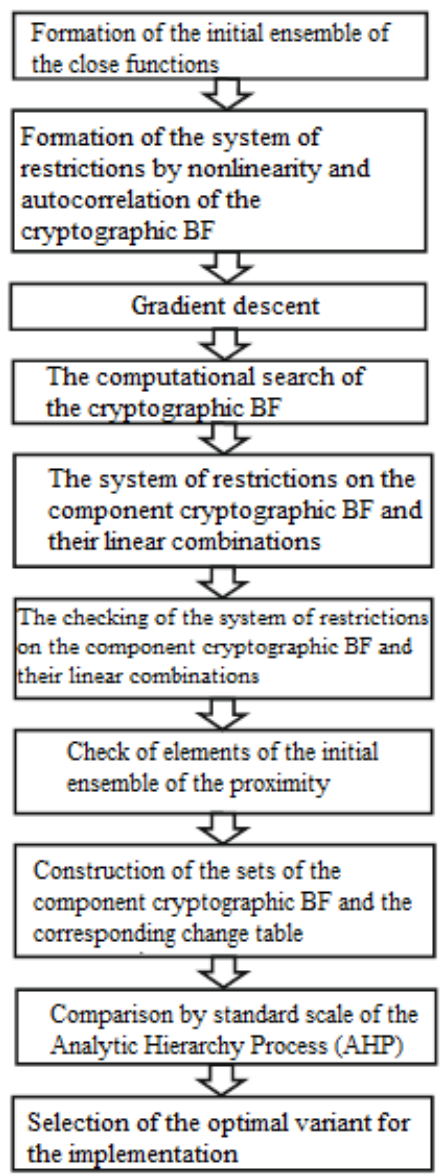

*Figure 2. The sequence of formation of the cryptographic functions in the Cognate-form representation*

As an example, we will study the procedure of formation of cryptographic functions using an alternative FR. Taking advantage of the algorithm of forming of BF of Cognate-FR [4,5,6] developed by me previously, and the method described in [5], we can offer the sequence of formation cryptographic function particularly in Cognate-FR (Fig. 2) which is described below.

Let us consider in more detail every item of this sequence.

1. The formation of the initial band of a Cognate functions. The initial ensemble $f_6$ close to the nominal BF $f_n$ is formed as a set of BF with have the single Cognate-proximity $C_{gn}$ to the nominal BF - $C_{gn} = \dfrac{1}{2^n}$. This gives the ensemble consisting of formally from the $2^{n+1}$ BF, but they will be subject to verification for acceptability complete with a nominal BF.
2. Formation of system of restrictions by nonlinearity and autocorrelation of the cryptographic BF. It used as a the initial information which defines the basic parameters of the calculation method of formation cryptographic BF using gradient search.
3. The procedures of computing research of cryptographic BF by the gradient descent method. The performed probabilistic search of BF use the method of the gradient descent with imposing the restrictions. The result is the randomly generated BF which satisfies the necessary value of nonlinearity and autocorrelation.
4. The constraints system for the cryptographic BF components and their linear combination is used as a the initial information which defines the basic parameters of the selection of the randomly stacked BF that satisfy the required values of the nonlinearity and autocorrelation.
5. The procedure of verification of perform of restrictions system on the components functions and their linear combinations formed by BF with the required values of nonlinearity and the autocorrelation are exposed for check for compliance with the requirements, meaning the suitability of use in combination with other BF.
6. Verification of elements of the initial ensemble of the cognate functions. Formation of the working ensemble of acceptable-cognate BF use strikeouts of the ensemble of elements which don't provide the real cognates.
7. Construction of the sets of the component cryptographic BF and the corresponding table of replacements.
8. Pairwise comparison according to standard scale of the hierarchies analysis method of the quality criteria and alternatives.

9. Election of the optimal variant for implementation. On the basis of a selected BF create the device which implements this embedded of transformation logic.

Thus Cognate-implementation of the Boolean functions differs from the classical one by performing additional steps:

- the formation of initial ensemble of acceptable options;
- the formation of a working ensemble of options after checking and curtailment of the elements of the original ensemble;
- the formation of plural quality assessment criteria of the implementation variants of BF;
- the formation of a plurality of the "close" alternatives;
- pairwise comparison according to standard scale of the hierarchy analysis method of the quality criteria and alternatives;
- choosing optimal variant for implementation.

CONCLUSIONS

The proposed in the article analysis by the example of Cognate-FR of BF indicates, that the use of alternative FR of BF when constructing cryptographic functions, algorithms and devices can significantly improve their parameters and properties. And their employment in the cryptographic means of protection permits to optimize process of logical design of the cryptographic protection devices and improve the safety performance of information and communication systems. Using the

Cognate-implementation provides the basis for a substantial reduction of hardware costs in the implementation of BF in the cryptographic devices. Therefore this area remains relevant scientifically and practically for scientists and developers of information security systems.